\documentclass[preprint]{aastex}
\newcommand{\bfm}[1]{{\mbox{\boldmath $#1$}}}
\def\ltsima{$\; \buildrel < \over \sim \;$}
\def\lsim{\lower.5ex\hbox{\ltsima}}
\def\gtsima{$\; \buildrel > \over \sim \;$}
\def\gsim{\lower.5ex\hbox{\gtsima}}
\def\Mpc{{\rm Mpc}}
\shortauthors{Szalay et al.}
\shorttitle{KL Power Spectrum Analysis of Early SDSS Data}
\begin{document}
\title{
KL Estimation of the Power Spectrum Parameters from the Angular 
Distribution of Galaxies in Early SDSS Data
}
\author{
Alexander S. Szalay\altaffilmark{9}, 
Bhuvnesh Jain\altaffilmark{9,14},
Takahiko Matsubara\altaffilmark{16},
Ryan Scranton\altaffilmark{2,3}, 
Michael S. Vogeley\altaffilmark{15},
Andrew Connolly\altaffilmark{4},
Scott Dodelson\altaffilmark{2,3},
Daniel Eisenstein\altaffilmark{5},
Joshua A. Frieman\altaffilmark{2,3},
James E. Gunn\altaffilmark{6}, 
Lam Hui\altaffilmark{8}, 
David Johnston\altaffilmark{2,3}, 
Stephen Kent\altaffilmark{3},
Martin Kerscher\altaffilmark{9}, 
Jon Loveday\altaffilmark{10}, 
Avery Meiksin\altaffilmark{18},
Vijay Narayanan\altaffilmark{6},
Robert C. Nichol\altaffilmark{11}, 
Liam O'Connell\altaffilmark{10},
Adrian Pope\altaffilmark{9}, 
Roman Scoccimarro\altaffilmark{7,12}, 
Ravi K. Sheth\altaffilmark{3}, 
Albert Stebbins\altaffilmark{3},
Michael A. Strauss\altaffilmark{6}, 
Istv\'an Szapudi\altaffilmark{13}, 
Max Tegmark\altaffilmark{14}, 
Idit Zehavi\altaffilmark{3},
James Annis\altaffilmark{3}, 
Neta Bahcall\altaffilmark{6},
Jon Brinkmann\altaffilmark{19},
Istv\'an Csabai\altaffilmark{20},
Masataka Fukugita\altaffilmark{23},
Greg Hennessy\altaffilmark{22}, 
David Hogg\altaffilmark{7},
Zeljko Ivezic\altaffilmark{6},
Gillian R. Knapp\altaffilmark{6}, 
Peter Z. Kunszt\altaffilmark{21},
Don Q. Lamb\altaffilmark{2},
Brian C. Lee\altaffilmark{3},
Robert H. Lupton\altaffilmark{6}, 
Jeffrey R. Munn\altaffilmark{17}, 
John Peoples\altaffilmark{3}, 
Jeffrey R. Pier\altaffilmark{17}, 
Constance Rockosi\altaffilmark{2}, 
David Schlegel\altaffilmark{6}, 
Christopher Stoughton\altaffilmark{3}, 
Douglas L. Tucker\altaffilmark{3},
Brian Yanny\altaffilmark{3}, 
Donald G. York\altaffilmark{2},
for the SDSS Collaboration
}

\altaffiltext{1}{Based on observations obtained with the 
	Sloan Digital Sky Survey}
\altaffiltext{2}{Astronomy and Astrophysics Department, 
        University of Chicago, Chicago, IL 60637, USA}
\altaffiltext{3}{Fermi National Accelerator Laboratory, 
        P.O. Box 500, Batavia, IL 60510, USA}
\altaffiltext{4}{University of Pittsburgh, Department of Physics and
        Astronomy, Pittsburgh, PA 15260, USA}
\altaffiltext{5} {Steward Observatory, The University of Arizona, 
        Tucson, AZ 85721, USA}
\altaffiltext{6}{Princeton University Observatory, 
        Princeton, NJ 08544, USA}
\altaffiltext{7}{Department of Physics, New York University, 
        New York, NY 10003}
\altaffiltext{8}{Department of Physics, Columbia University, 
        New York, NY 10027, USA}
\altaffiltext{9}{Department of Physics and Astronomy, The Johns Hopkins 
        University, Baltimore, MD 21218, USA}
\altaffiltext{10}{Sussex Astronomy Centre, University of Sussex, 
        Falmer, Brighton BN1 9QJ, UK}
\altaffiltext{11}{Department of Physics, Carnegie Mellon University, 
        Pittsburgh, PA 15213, USA}
\altaffiltext{12}{Institute for Advanced Study, School of Natural Sciences,
        Princeton, NJ 08540, USA}
\altaffiltext{13}{Institute for Astronomy, University of Hawaii, 
        Honolulu, HI 96822, USA}
\altaffiltext{14}{Department of Physics, University of Pennsylvania,
        Philadelphia, PA 19101, USA}
\altaffiltext{15}{Department of Physics, Drexel University, 
        Philadelphia, PA 19104, USA}
\altaffiltext{16}{Department of Physics and Astrophysics, 
        Nagoya University, Chikusa, Nagoya, 464-8602, Japan}
\altaffiltext{17} {U.S. Naval Observatory, Flagstaff Station, P.O. Box 1149, 
	Flagstaff, AZ 86002-1149}
\altaffiltext{18} {Institute for Astronomy, University of Edinburgh, 
	Blackford Hill, Edinburgh EH9 3HJ, UK}
\altaffiltext{19} {Apache Point Observatory, P.O. Box 59, Sunspot, NM 88349}
\altaffiltext{20} {Department of Physcs, E\"otv\"os University, Budapest, 
	H-1117, Hungary}
\altaffiltext{21} {CERN, IT Division, 1211 Geneva 23, Switzerland}
\altaffiltext{22} {US Naval Observatory, Washington DC 20392-5420, USA}
\altaffiltext{23} {Institute for Cosmic Ray Research, University of Tokyo, Kashiwa, Chiba 277-8582, Japan}

\begin{abstract}
We present measurements of parameters of the 3-dimensional power
spectrum of galaxy clustering from 222 square degrees of early imaging
data in the Sloan Digital Sky Survey.  The projected galaxy
distribution on the sky is expanded over a set of Karhunen-Lo\`eve
eigenfunctions, which optimize the signal-to-noise ratio in our
analysis.  A maximum likelihood analysis is used to estimate
parameters that set the shape and amplitude of the 3-dimensional power
spectrum. Our best estimates are $\Gamma=0.188\pm 0.04$ and
$\sigma_{8L} = 0.915 \pm 0.06$ (statistical errors only), for a flat
universe with a cosmological constant. We demonstrate that our
measurements contain signal from scales at or beyond the peak of the
3D power spectrum. We discuss how the results scale with
systematic uncertainties, like the radial selection function. We find
that the central values satisfy the analytically estimated scaling
relation.  We have also explored the effects of evolutionary
corrections, various truncations of the KL basis, seeing, sample size
and limiting magnitude. We find that the impact of most of these
uncertainties stay within the 2$\sigma$ uncertainties of our fiducial
result.
\end{abstract}


\section{Introduction}

Galaxy surveys have been widely used to map large-scale structure in
the universe. While redshift surveys map the full 3-dimensional
distribution of nearby galaxies, imaging surveys that map the galaxy
distribution on the sky probe higher redshifts and sample a much
larger number of galaxies. The APM survey is the largest existing
imaging survey and has been used to estimate the 3-dimensional power
spectrum of galaxy clustering (Baugh \& Efstathiou 1994; Dodelson \&
Gaztanaga 2000; Eisenstein \& Zaldarriaga 2001; Efstathiou \& Moody
2001).

To estimate the 3-dimensional power spectrum from an angular survey
requires de-projection of the data. In the absence of any redshift
information this is done using Limber's equation with estimates of the
redshift distribution based on the magnitude limit of the survey
(Limber 1953; Peebles 1980). The 3-dimensional power spectrum
estimates from the APM survey employ this technique.  

This paper is part of the first results (Scranton et al. 2001,
Connolly et al. 2001, Dodelson et al. 2001, Tegmark et al. 2001,
Zehavi et al.  2001) on large scale clustering of galaxies from the
Early Data Release (EDR) of the Sloan Digital Sky Survey (SDSS). The
EDR data (Stoughton et al. 2001) cover approximately 600 square
degrees, roughly 6\% of the final sky coverage of the survey, mostly 
in two equatorial slices.  The data set contains over 8 million galaxies
in 5 color photometry with limiting magnitude $r^*\approx 22.5$
(detection limit of $5:1$ signal-to-noise ratio) (Fukugita et al 1996,
Gunn et al 1998, Lupton et al 2001, Stoughton et al 2001). An
extensive effort has been carried out to understand the systematic and
statistical issues affecting the various measures of angular
clustering in this data set (Scranton et al. 2001). In order to enable
fair comparisons between different statistical techniques used, we
have selected a common subset of the EDR data to be used for the
current set of papers, called EDR-P. This area covers about 222 square
degrees.

This paper focuses on the measurement of parameters from second order
statistics using the imaging data in the EDR-P data set.  Here we
present results for the shape and normalization of the 3-dimensional
power spectrum.  Section 2 provides the theoretical framework of
Karhunen-Lo\`eve eigenfunction expansions that is used to estimate the
parameters of the power spectrum.  In section 3 we describe the data
set and the details of the analysis. In Section 4 we apply the KL
method to the data to estimate parameters of the 3-dimensional power
spectrum. We conclude in Section 5 with comparison of our parameter
estimates with results from other SDSS analyses, other redshift
surveys, and other cosmological constraints.

\section{Formalism}

Limber's equation is used to predict the angular
clustering for an input cosmological model. Basic parameters of the
cosmology -- the matter and vacuum energy density and dark matter
constituent -- are taken as fixed, and the shape and normalization of
the galaxy power spectrum are fitted using Maximum-Likelihood
estimation from the coefficients of an eigenfunction expansion of the
observed data.  The following subsections present the formalism for
this approach.  We consider only models with a flat geometry. Our
fiducial model is $\Omega_m=0.3$, $\Omega_\Lambda=0.7$, in agreement with
recent constrains from CMB fluctuations (see, e.g., Netterfield et 
al. 2001, Lee et al. 2001, Halverson et al. 2001)

\subsection{Limber's Equation for the Angular Correlation Function}

Limber's equation expresses the angular correlation function in terms
of the 3-dimensional power spectrum of the galaxy distribution
$P(\chi, k)$ ($P(k)$ at the epoch corresponding to comoving distance
$\chi$) as
\begin{equation}
    \omega(\theta)= 4 \pi^2 \int_0^{\chi_H} d \chi \ W(\chi)^2\
	\int_0^\infty dk\, k\, P(\chi, k)\, 
	J_0\left[k r(\chi)\theta\right] \;.
\label{wtheta}
\end{equation}
where $W(\chi)=n(z)H(z)/c$ denotes the radial distribution of galaxies
in the sample and $\chi_H$ is the distance to the horizon.
Here our notation is such that the
unperturbed Robertson-Walker metric is
\begin{equation}
    ds^2=a^2(\tau)\left( -d\tau^2+ d\chi^2+
	r^2(d\theta^2+\sin^2 \theta d\phi^2) \right)  ,
\label{metric}
\end{equation}
where $\tau$ is conformal time, and $a(\tau)$ is the expansion scale factor.
Thus, the comoving angular
diameter distance $r(\chi)$ is
\begin{eqnarray}
    r(\chi)=\sin_K\chi \equiv
    \left\{ 
	\begin{array}{ll} K^{-1/2}\sin K^{1/2}\chi,\ K>0\\
	    \chi, \ K=0\\(-K)^{-1/2}\sinh (-K)^{1/2}\chi,\ K<0 
	\end{array}
    \right.
\label{rchi}
\end{eqnarray}
where $K$ is the spatial curvature given by
$K=-H_0^2(1-\Omega_m-\Omega_\Lambda)$ with $H_0$ being the Hubble
parameter today.

\subsection{Expansion of the Galaxy Distribution 
	into Karhunen-Lo\`eve Eigenfunctions}

The Karhunen-Lo\`eve (KL) eigenfunctions (Karhunen 1947, Lo\`eve 1948)
provide a basis set in which the distribution of galaxies can be
expanded. These eigenfunctions are computed for a given survey
geometry and fiducial model of the power spectrum. For a Gaussian
galaxy distribution, the KL eigenfunctions provide optimal estimates
of model parameters, i.e. the resulting error bars are given by the
inverse of the Fisher matrix for the parameters Vogeley \& Szalay
1996).  This is achieved by finding the orthonormal set of
eigenfunctions that optimally balance the ideal of Fourier modes with
the finite and peculiar geometry and selection function of a real
survey. In this section we present the formalism for the KL analysis
following the notation of Vogeley \& Szalay (1996) who introduced this
approach to galaxy clustering. The KL method has been applied to the
Las Campanas redshift survey by Matsubara, Szalay \& Landy (2000) and
to the PSCz survey by Hamilton, Tegmark \& Padmanabhan (2001).

The angular distribution of galaxies is pixelized by dividing the
survey area into a set of $N$ cells. The data vector can then be
defined as
\begin{equation}
    d_i=n_i^{-1/2}(m_i-n_i)
\label{datavec}
\end{equation}
where $m_i$ is the number of galaxies in the $i$-th cell, $n_i=\langle
m_i\rangle$ is the expected number of galaxies and the factor
$n_i^{-1/2}$ is included to whiten the shot noise as explained below.
The data vector $\bfm d$ is expanded into the set of KL eigenfunctions
$\bfm \Psi_n$ as
\begin{equation}
    \bfm{d} = \sum_n B_n \bfm{\Psi}_n.
\label{eigenexp}
\end{equation}
The eigenfunctions $\bfm \Psi_n$ are obtained by solving the
eigenvalue problem (Vogeley \& Szalay 1996):
\begin{equation}
   \bfm{R} \bfm{\Psi}_n = \lambda_n \bfm{\Psi}_n,
\label{evalue}
\end{equation}
where $\lambda_n = \langle B_n^{2} \rangle$ and
\begin{equation}
   R_{ij} = \langle d_i d_j \rangle = 
   n_i^{1/2} n_j^{1/2} \omega_{ij} + \delta_{ij} \ .
\label{corrmatrix}
\end{equation}
The second term is the whitened shot noise correlation matrix.  The
correlation matrix $\bfm{R}$ is computed for a fiducial model using
the cell-averaged angular correlation function
\begin{equation}
    \omega_{ij}\equiv \frac{1}{V_i V_j}\int \int d^2\theta_i 
	\ d^2\theta_j \ \omega(|\bfm{\theta}_i-\bfm{\theta}_j|) \ ,
\label{cellave}
\end{equation}
where the integral extends over the areas of the $i$-th and $j$-th
cells, and $V_i$ and $V_j$ are the corresponding cell areas.  Forming
the eigenmodes $\bfm{\Psi}_n$ requires assuming an a priori model for
$\omega(\theta)$ but, as discussed by Vogeley \& Szalay (1996), this
choice does not bias the estimated parameters below.

The KL eigenmodes defined above satisfy the conditions of
orthonormality $\bfm{\Psi}_n \cdot \bfm{\Psi}_m = \delta_{nm}$, and
statistical orthogonality, $\langle B_n B_m \rangle = \langle
B_n^{\,2} \rangle \delta_{nm}$. Further, they sort the data in
decreasing signal-to-noise ratio if they are ordered by the corresponding
eigenvalues (Vogeley \& Szalay 1996). What this means in the
measurement of model parameters will be clarified below.

The KL expansion is used to estimate model parameters by computing the
covariance matrix $\bfm{C}$ of the KL coefficients. We use the first
$N_{mode}$ of the KL eigenmodes and choose to parameterize the model
by the linear amplitude (r.m.s. of density field) at $8h^{-1}$Mpc
$\sigma_{8L}$ and shape $\Gamma$ for a CDM-like power spectrum
($\Gamma\approx \Omega h$ -- see Efstathiou, Bond, \& White 1992;
Peacock \& Dodds 1994).  The theoretical covariance matrix is then
given by
\begin{equation}
    C_{mn} = \langle B_m B_n \rangle_{\rm model}
	= \bfm{\Psi}_m^T \bfm{R}_{\rm model} \bfm{\Psi}_m \ .
\label{covariance}
\end{equation}
$\bfm{R}_{\rm model}$ is computed by using the given power spectrum
and computing $\omega_{ij}$ for a given cosmology using equation
\ref{wtheta}. This includes evolution of the power spectrum with
comoving distance from the observer, $P(\chi,k)$, as specified by the
fiducial model.  Note that $\bfm{C}$ is not diagonal in general unless
the model parameters are identical to those of the fiducial model used
for computing the fixed set of eigenmodes $\bfm{\Psi}$.  We use an
unbiased cluster normalized CDM model, with $\Gamma=0.25$,
$\sigma_{8L}=1$ for the galaxy power spectrum and an
$\Omega_\Lambda=0.7$, $\Omega_m=0.3$ cosmology for our fiducial model.
Because the model is unbiased, this assumes that the evolution of
galaxy clustering is identical to the evolution of mass clustering
over the range of redshifts probed by this sample.  As discussed
below, the final parameter estimates yield $\sigma_{8L}\approx 1$ for the
galaxies, so this assumption is not unreasonable. 

If the galaxy density field is Gaussian then the likelihood function
of the data is a multivariate Gaussian given by
\begin{equation}
   {\cal L} = (2\pi)^{-N_{mode}/2} |\det\bfm{C}|^{-1/2}
   \exp\left[-\frac12 \bfm{B}^T \bfm{C}^{\,-1} \bfm{B}\right]\ .
\label{like}
\end{equation}
Maximizing the log-likelihood yields the best fit model parameters.
We have tested our KL package on simulations by Cole et al. (1998).
The input cosmological parameters were well recovered.

Advantages of this approach (as discussed by Vogeley \& Szalay 1996)
are that (1) it linearly transforms the data into a basis of nearly
uncorrelated modes (exactly uncorrelated in the case of the fiducial
model), which makes hypothesis testing much easier because the
correlation matrices are nearly diagonal, (2) the modes are sorted by
signal-to-noise ratio, so a truncation of the transformed data set
maintains maximum fidelity of the original data, and (3) the
covariance matrices of the transformed data depends on second moments
only.  In contrast, when using quadratic estimators, one needs to deal
with substantial covariance matrices of the density field, which
require knowledge of third and fourth order correlations.

\section{Results from SDSS Early Data}

\subsection{Selection of the Data}

The data are from the EDR-P, a subset of the SDSS Early Data Release
augmented with a Bayesian star/galaxy separation method producing
galaxy probabilities for each object (Scranton et al 2001).  The
separation method and extensive tests of the method against systematic
errors both external (seeing variations, dust extinction, stellar
contamination, bright stars, and sky brightness) and internal (uniform
photometric response and calibration, Limber magnitude scaling and
deblending efficiency) are described by Scranton et al. (2001).
Adopting the convention of the other papers using the EDR-P, we split
the data into unit magnitude bins based on each object's model
magnitude in $r^*$ (York et al., 2000).  We use three of the magnitude
bins adopted by the other papers: $18 < r^* <19$, $19 < r^* < 20$, and
$20 < r^* < 21$.  We do not analyze the $21<r^*<22$ magnitude bin of
the EDR-P to avoid dealing with the complex small-scale angular mask
(see Scranton et al. 2001), which is important only for this very
deepest subsample.

The angular region covered by the EDR-P sample is a narrow equatorial
stripe $2.5^\circ$ degrees in declination and running from
$9^{h}44^{m}59^{s}$ to $15^{h}37^{m}23^{s}$ in right ascension (J2000),
which yields a solid angle of approximately 222 square degrees. For each
magnitude bin, we pixelize the data area using pixels 0.5 degrees on a
side.  The number of galaxies in a given pixel is the sum of the
galaxy probabilities for all the objects in the pixel for a given
magnitude bin.  Calculating the mean number of galaxies in all the
pixels yields the expected number of galaxies per pixel.

\subsection{The Shape of the Assumed Redshift Distribution}

\begin{figure}[t!]
\vspace{8cm}
\caption{
Redshift distributions assumed for the magnitude bins
$18<r^*<19$, $19<r^*<20$ and $20<r^*<21$ shown by solid lines
corresponding to increasing values of the median redshift. The two
dashed curves show the uncertainty in the estimated redshift
distribution for the $20<r^*<21$ bin.
}
\includegraphics{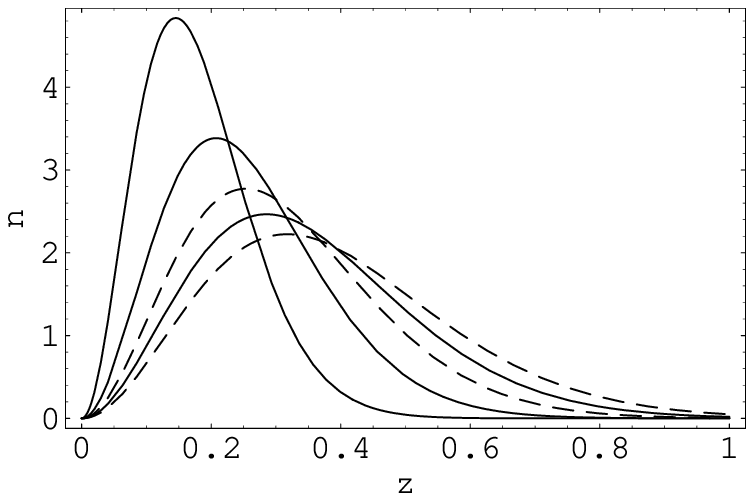}
\label{fig_sf}
\end{figure}

The redshift distribution of the galaxies was approximated by
$dn/dz\, \propto z^2 \exp(-(z/z_0)^{1.5})$, with the median
redshift $z_m=1.412 z_0$.  We use median redshifts $z_m=0.17, 0.24,
0.33$ for the three magnitude bins, $18< r^*< 19$,
$19< r^*<20$ and $20<r^*<21$, respectively.  Dodelson et
al (2001) give a detailed description of how the redshift distribution
was obtained using the CNOC2 survey (Lin et al. 1999) and corrected
for differences from the SDSS magnitude system.  Figure \ref{fig_sf}
shows the redshift distributions, normalized to have unit integral
over redshift. The dashed curves show estimates of the uncertainty in
the redshift distribution for the $20<r^*<21$ bin, based on the
standard deviations in $z_m$ derived by Dodelson et al. (2001).  We
will use these distributions to estimate the sensitivity of the power
spectrum parameters we obtain to uncertainty in the redshift
distribution.

\subsection{Building the KL Basis}

We use the geometry of the $5 \times 175$ pixel map to build our KL
basis, using the fiducial model. We precompute the angular correlation
function $w(\theta)$, store these values in a table, and use
interpolation to calculate its values. For close-by pairs of cells we
use a direct numerical integration of Equation 8.  For distant cells
we use the separation between the cell centers.  For each relative
cell-pair we use hash codes ( a single integer used as index to an
array) to uniquely define the relative geometry and compute similar
configurations only once and store those values in a table for reuse.

\begin{figure}[t!]
\vspace{10.5cm}
\caption{
The eigenvalues of the correlation matrix are plotted vs.  the mode
number of the KL eigenmodes for the magnitude bin $20<r^*<21$ (upper
panel). These are ordered by decreasing signal-to-noise, but the
normalization is arbitrary. The transition in slope at around mode 40
approximately corresponds to the aspect ratio of our $5\times 175$
pixel map, thus marking the transition from purely ``longitudinal'' or
1-dimensional to genuine 2-dimensional modes. The lower panel shows
the effective wavenumber for a given mode, corresponding to the peak
of the power spectrum of the mode, where the angular wavelength was
converted to transverse length at the mode of our redshift
distribution, using our fiducial cosmology.
}
\includegraphics{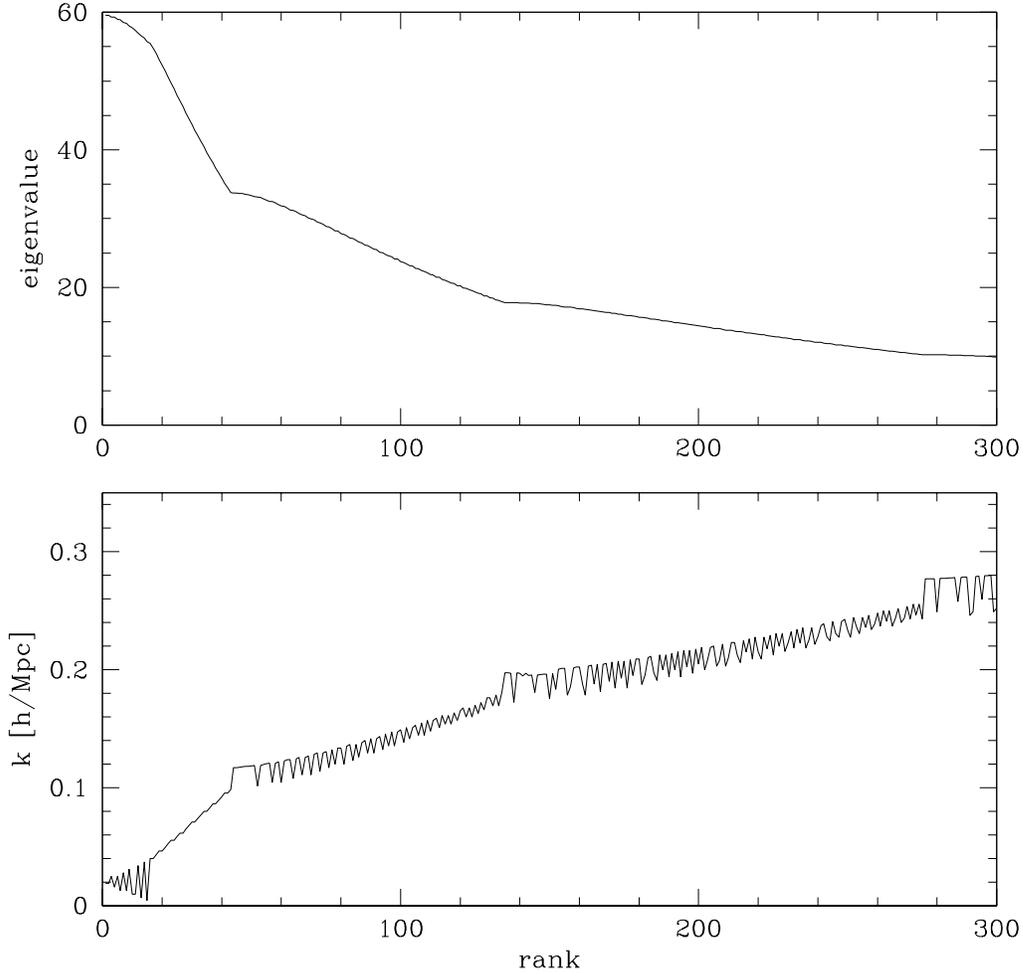}
\label{fig_eval}
\end{figure}

\begin{figure}[t!]
\vspace{6cm}
\caption{
Selected KL modes (mode numbers, or ranks 1,2,4,16 and 300) are shown for the
$2.5^{\circ} \times 90^{\circ}$ geometry of the data stripe. The
lowest panel shows mode 300, showing the smallest length scales,
corresponding to about 0.5$^{\circ}$, used in the parameter
estimation.
}
\includegraphics{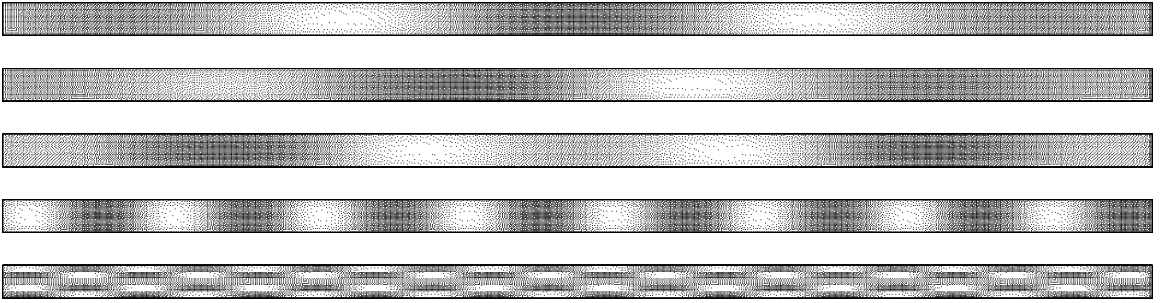}
\label{fig_modes}
\end{figure}

\begin{figure}[t!]
\vspace{6cm}
\caption{
The input and reconstructed pixellized density field of the data. The
upper panel shows the data binned in pixels 0.5$^{\circ}$ on a
side. The middle panel shows the reconstructed density using the first
300 KL modes. The bottom panel shows the residuals.
}
\includegraphics{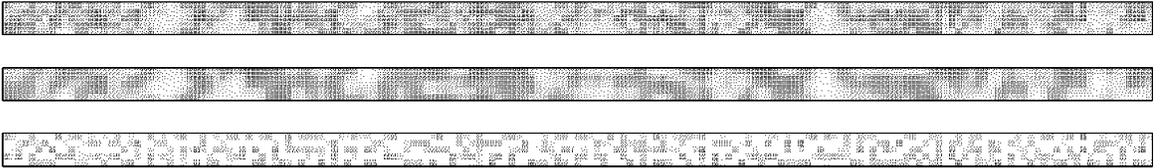}
\label{fig_recon}
\end{figure}

We use a uniform expected surface density of galaxies for our noise
estimation. We whiten this noise as described in Section 2.  We
compute the first 300 from a total of 875 modes. The eigenvalue
spectrum is shown in Figure \ref{fig_eval}. Selected modes are
displayed in Figure \ref{fig_modes}.  Below we examine the sensitivity
of our results to this truncation, to ensure that we are safely in the
regime where linear theory is applicable and non-linear corrections
can be ignored, and find that $N_{mode}=250$ is an appropriate cutoff
for our power spectrum analyses.  Because higher numbered modes
primarily sample high frequencies, this truncation results in a
smoothing of the galaxy surface density. The top image of Figure
\ref{fig_recon} displays the pixel values corresponding to the $d_i$
vector.

Figure \ref{fig_modesxy} shows how the modes are distributed in
2-dimensional k-space. Since the survey geometry leads to an elongated
window in k-space, the KL modes are also elongated as no mode can be
narrower than this window. The KL modes are orthogonal and represent
an approximate dense packing of the allowed region in k-space,
starting at the origin and proceeding outwards in shells.
Representative modes shown in the figure as ellipses (with ranks just
below 300 shown by dotted curves, and just below 250 shown by the
solid curves) illustrate the angular distribution and elongation of
the KL modes. It should be noted that transverse modes (oriented along
the y-axis) mix a wider range of wavenumber amplitudes than
longitudinal modes (oriented along the x-axis). Hence the longitudinal
modes provide sharper probes of the power spectrum at a given
wavenumber amplitude.

\begin{figure}[t!]
\vspace{10.5cm}
\caption{
Histogram of the first 300 KL coefficients for the $20<r^*<21$ sample,
normalized as $b_n = B_n/\sqrt{\lambda_n}$. These should have a normal
Gaussian distribution (shown by the solid curve) for a Gaussian
density field.  The agreement of the measured histogram with the
Gaussian curve demonstrates that the number of KL modes has been
chosen appropriately -- there are no features arising from small-scale
nonlinearity/non-Gaussianity.
}
\includegraphics{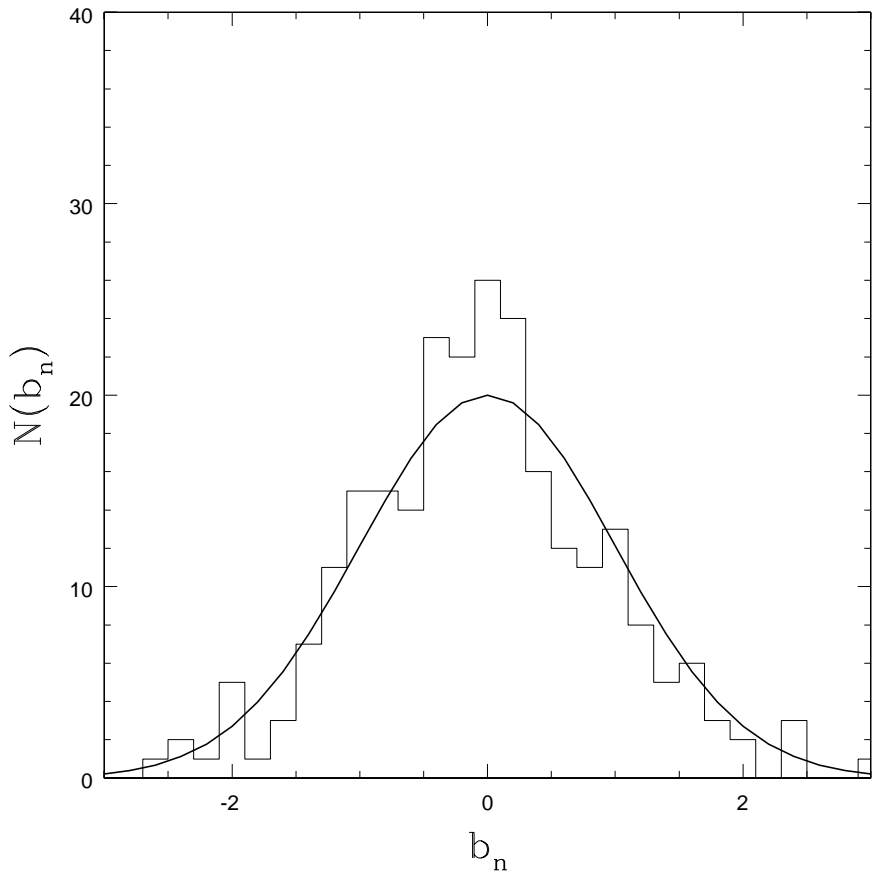}
\label{fig_hist}
\end{figure}

We transform the data into the KL coefficients $B_n$ by computing the
scalar product of the $n$th mode with the $d_i$ vector, $B_n= \bfm{d}
\cdot \bfm{\Psi}_n$.  Then we create the normalized KL-coefficients
\begin{equation}
    b_n = B_n/\sqrt{\lambda_n}.
\end{equation}
\label{bn}
These are expected to have a normal Gaussian distribution, if our
truncation avoids the modes where non-linear contributions may be
important.  The amplitude distribution of the first 300 $b_n$ is shown
in Figure \ref{fig_hist}. The distribution of these coefficients is
rather close to a Gaussian, although there is a slight asymmetry in
the distribution.

The 300 coefficients $b_n$ are used to reconstruct the smoothed
density. This is shown in comparison to the original pixelized data
and the residual on Figure \ref{fig_recon}. Note that the residual sky
map contains only very high frequencies, close to the pixel level,
thus most of the information on large-scale clustering is included in
the first 300 modes.

\section{Results from the likelihood analysis} 

\subsection{Our fiducial case: $\Omega_\Lambda=0.7$, $\Omega_m=0.3$}

We use the vector of KL coefficients $b_n$ to compute the likelihood.
Since the $b_n$ are normalized, we also need to transform the original
correlation matrix to the correlation of the $b_n$.  In this
transformed space, if we compute the correlation matrix with the
fiducial parameter values that were used to construct our basis, then
the transformed correlation matrix will be the identity matrix. This
transformation involves a projection, a rotation and a renormalization
by $\sqrt\lambda_n$ of the original correlation matrices for each
model to be tested. We find it necessary to use only 250 of the
possible 875 modes, thus the transformed correlation matrix (to be
inverted) is only $250\times 250$, instead of the full $875\times875$.

\begin{figure}[t!]
\vspace{10.5cm}
\caption{
Likelihood contours (plotted at 1, 2, and $3\sigma$) for $\sigma_{8L}$
and the shape parameter $\Gamma$ for three magnitude bins as indicated
in the panels. These assume an $\Omega_m=0.3, \Omega_\Lambda=0.7$
cosmology. The redshift distribution dn/dz$\ \propto z^2
e^{-(z/z_0)^{1.5}}\ $ with median redshifts $z_m = 0.17, 0.24, 0.33$
is used for the three panels respectively ($z_m=1.412 z_0$). The KL
expansion was truncated at 300 modes in the upper three panels. To
ensure that only the linear modes are used, in the lower panels the
expansion was truncated at 60, 150 and 250 modes (from left to right).
The panel in the lower right is our best fit.
}
\includegraphics{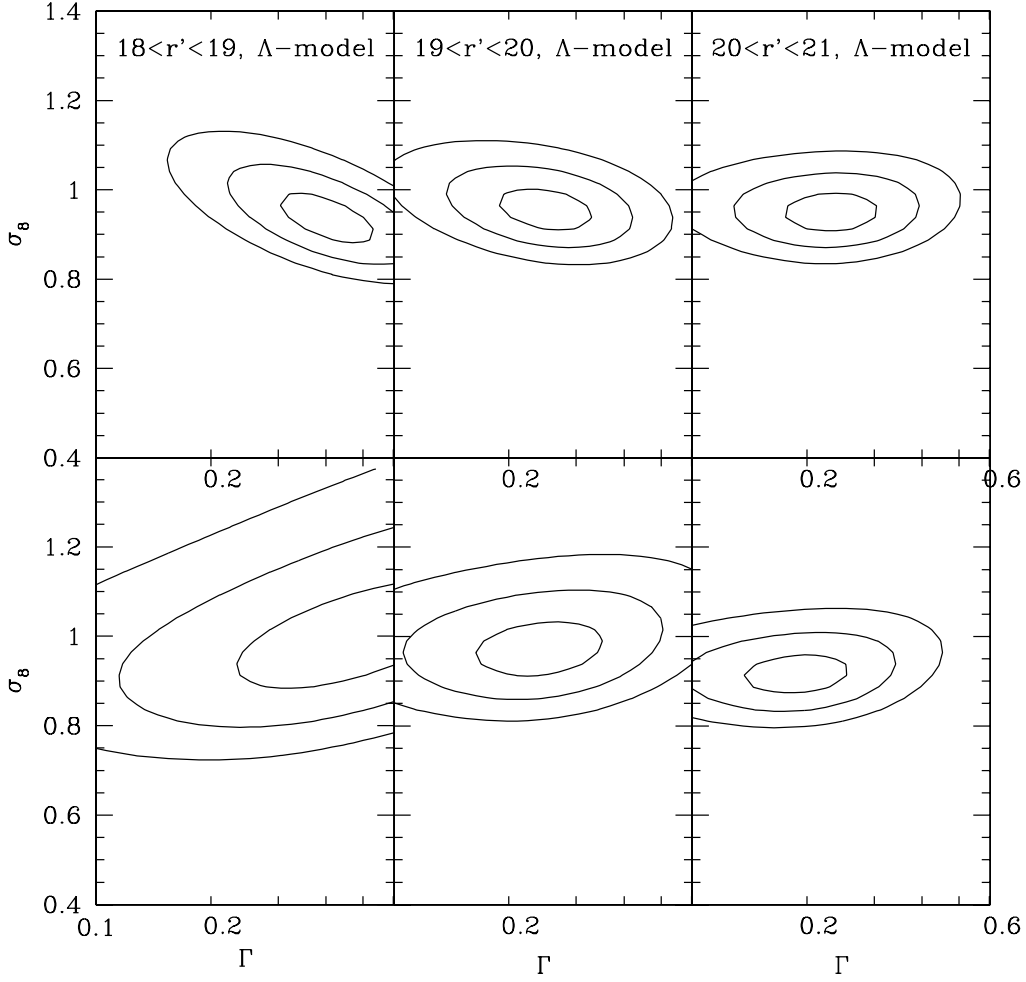}
\label{fig_lambda}
\end{figure}

First we present the likelihood contours for our fiducial cosmology.
We fix $\Omega_m=0.3$ and $\Omega_\Lambda=0.7$, and vary the values of
$\Gamma$ and $\sigma_{8L}$. In the latter quantity the subscript $L$
means that this is the `linear' $\sigma_{8L}$, reflecting the
amplitude of the power spectrum without any non-linear corrections.
Note that the correlation matrix $\bfm{C}$ computed for each model
includes evolution of the power spectrum predicted by that model
(through equation \ref{wtheta}), thus $\sigma_{8L}$ is an estimate of
the linear clustering amplitude at the present epoch, not the
amplitude at the effective redshift of each galaxy sample.  Figure
\ref{fig_lambda} shows the 1, 2, and $3\sigma$ likelihood contours
\footnote[1]{Note carefully the meaning of likelihood contours such as
those in Figure \ref{fig_lambda} and below: The $1\sigma$ contour, for
example, is drawn at $\Delta \chi^2=1$ from the maximum likelihood and
therefore can be used to marginalize ``by eye'' to obtain the
$1\sigma$ (68\% confidence interval) limits on each parameter
separately by examining the height and width of the error
ellipse. However, because there are two degrees of freedom to be fit,
this ``$1\sigma$'' contour encloses a smaller region than that which
includes 68\% of the bivariate likelihood and likewise for the 2 and
$3\sigma$ contours.  In other words, a point in the $\Gamma,
\sigma_{8L}$ plane just outside of the ``$2\sigma$'' contour (drawn at
$\Delta \chi^2=4$) is {\it not} ruled out at the 95\% confidence level
(see, e.g., Press et al. 1992).}  for the three magnitude bins
$18<r^*<19, 19<r^*<20, 20<r^*<21$. In the projection of clustering onto
the sky we have assumed median redshifts $z_m$=0.17, 0.24 and 0.33,
respectively, as described above.

The upper three panels in Figure \ref{fig_lambda} used 300 modes for
the likelihood analysis. In the lower three panels we use fewer modes
in order to restrict the analysis to the linear regime of clustering
as discussed below. Thus we use 60, 150 and 250 modes for the three
magnitude bins $18<r^*<19, 19<r^*<20, 20<r^*<21$, respectively. As
evident in the figure, the errors on the parameters in the brightest
bin are very large with just 60 modes, but in the two fainter
magnitude bins we still get interesting constraints.  As discussed by
Dodelson et al (2001), with increasing depth the data at a given
angular scale have smaller clustering amplitude, thus allowing us to
use a larger dynamic range for parameter estimation. Further,  the
number of galaxies is larger, resulting in lower shot noise. For the
faintest bin used in our analysis, we find
\begin{equation}
\Gamma=0.188\pm0.04, \qquad \sigma_{8L}=0.915\pm0.06 
        \qquad {\rm for}\ 20<r^*<21.
\label{values}
\end{equation}
Quoted errors on each parameter are $1\sigma$ (68\% confidence region
marginalized over the other parameter).  Again, note that these are
fits for the linear power spectrum extrapolated to $z=0$; these are
not estimates of the parameters at $z_m$ of each sample.  These values
are statistically independent from one another, since there was no
overlap between the samples (although there is some cosmic covariance
because the volumes sampled by the different galaxies do overlap).
The variation in the parameter values between the deeper two bins is
rather mild, while in the brightest bin the value of $\Gamma$ is high
in comparison. The same variation with sample depth of the estimated
parameters is seen in the angular power spectrum coefficients (Tegmark
et al. 2001). Note also that cosmic variance is largest for the
brightest bin, which has the smallest volume and total number of
galaxies, thus the uncertainties on parameters for this nearest
subsample are relatively large.

Perhaps more important for the brightest sample, nonlinear effects
become more prominent as smaller length scales and lower redshifts are
probed, leading to a power spectrum shape with more small scale power.
Because our parameter fits are those of a {\it linear} power spectrum,
increased sensitivity to smaller more nonlinearly evolved scales, on
which the power per mode exceeds the linear prediction, will tend to
drag the fits toward larger $\Gamma$. 

An important feature of the fitted parameters is how their covariance
changes with the depth of the samples.  In the brightest bin the two
parameters are correlated, as manifested by the tilt in the
probability contours. This means that we cannot distinguish between a
left-right shift in the power spectrum (the effect of changing
$\Gamma$) and an up-down scaling (due to a change in
$\sigma_{8L}$). This has been the case with most angular inversions of
the power spectrum to-date, and reflects the fact that previous
relatively shallower data sets only sampled the falling, monotonic
part of the power spectrum, shortward of the turnover.

\begin{figure}[t!]
\vspace{6.5truein}
\caption{ 
The peak wavenumber corresponding to the first 300 KL modes for the
$20<r^*<21$ magnitude bin. The modes are numbered by their eigenvalues
(Figure \ref{fig_eval}). The modes 1-100 are shown as black dots, the
modes 101-200 are stars, the modes 201-250 are x's and 251-300 are
open circles. For some of the modes we also show the contours of the power
spectrum corresponding to the given mode.  The elongation of
the modes in $k$-space is due to the extreme aspect ratio of our
geometry.  The peak of the power spectrum is at about $k = 0.02
h/\Mpc$, within which there are six modes with high signal-to-noise. 
}
\includegraphics{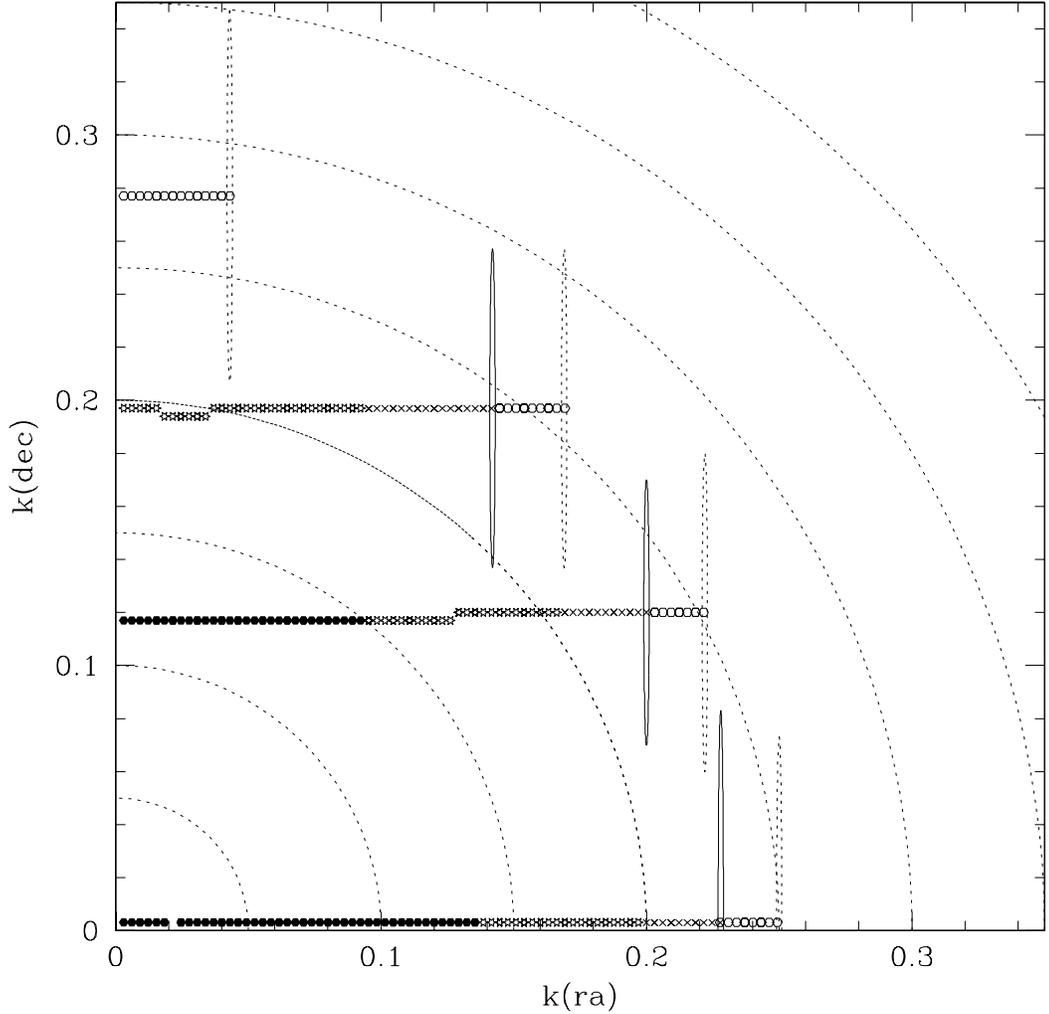}
\label{fig_modesxy}
\end{figure}

In a data set that is sufficiently deep to sample both sides of the
power spectrum peak efficiently, the two-parameter power spectrum
parameters become uncorrelated --- the covariance aligns with the
axes. This is exactly the case with our faintest sample. As shown by
Figure \ref{fig_modesxy}, we measure the power spectrum on both sides
of the peak! The transverse scale of our slice is quite large: in this
faintest bin it is well over a gigaparsec. The accuracy and the
statistical weight of the contributions coming from longward of the
peak is determined by the number of independent modes with a
wavelength longer than the peak. This fact shows the importance of
well-calibrated, wide area photometric surveys, such as the SDSS.

\begin{figure}[t!]
\vspace{10.5cm}
\caption{ 
Likelihood contours for $\sigma_{8L}$ and $\Gamma$ for various
truncations of the KL basis, using the magnitude bin $20<r^*<21$.  As
the number of modes is cut, we see that the likelihood contours become
broader, as the information used in the fitting process is reduced,
and a positive correlation between $\sigma_{8L}$ and $\Gamma$ appears,
since the low order modes mostly probe the rising slope of $P(k)$.
The central values of $\sigma_{8L}$ and $\Gamma$ are nearly identical
for 200 or 250 modes (middle panel) but these parameters are larger
for 300 modes, with no significant increase in precision. These extra
50 modes start to probe power in the mildly nonlinear regime (see
figure \ref{fig_modesxy}), yielding a shift in the estimated
parameters in the expected sense.  The figure shows that truncating at
250 modes provides small error bars and avoids leakage from nonlinear
scales at the same time.
}
\includegraphics{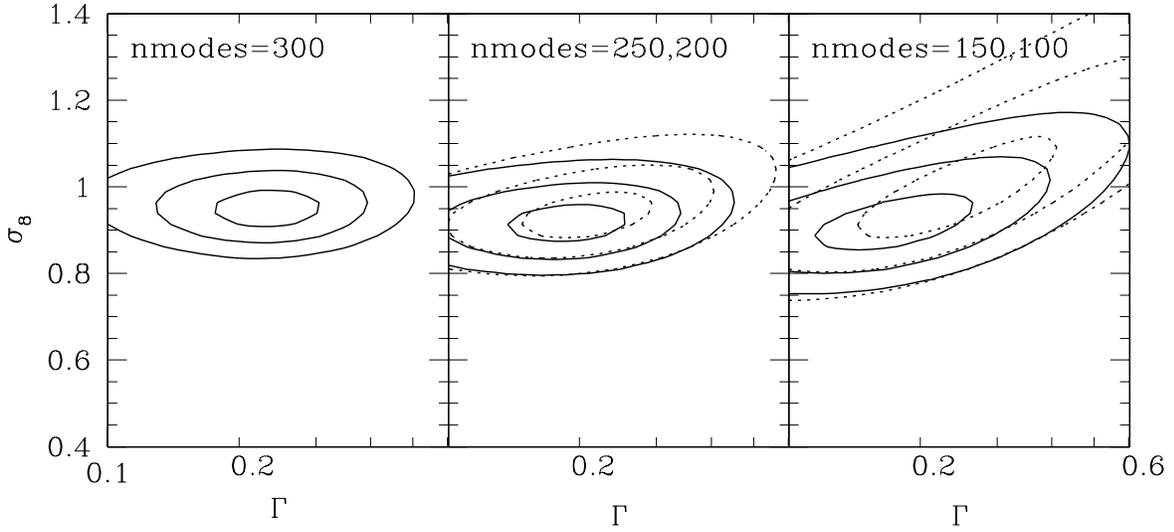}
\label{fig_trunc}
\end{figure}

Figure \ref{fig_trunc} illustrates the effect of varying the number of
KL modes used in the parameter fitting and justifies our choice of
$N_{mode}=250$ as the appropriate cutoff for the $20<r^*<21$
sample. As we increase the number of modes, the error contours shrink,
but for too large a number of modes we admit signal from nonlinear
scales. Here we see that the fitted parameters are stable and the
uncertainties decrease as we go from $N_{mode}=200$ to 250, but that
there is a slight bias toward larger $\Gamma$ as we go to 300 modes
because those extra 50 modes include some nonlinear power. This is
evident from Figure \ref{fig_modesxy} which shows the peak wavenumbers
for these modes. The ellipses of modes with rank 250-300 lie at
constant wavenumber $k\simeq 0.28$ along the y-axis and extend out to
$k\simeq 0.35$. Thus dropping these modes restricts our maximum
wavenumber to $k\simeq 0.25$, with most of the signal coming from
wavenumbers below $k = 0.2$.

At the other extreme, if we remove a large number of modes (see right
panel of Figure \ref{fig_trunc}), the uncertainties becomes
unacceptably large because we are throwing out much useful information
about the clustering. Note also in this last panel that the covariance
of parameters tilts in opposite fashion to the results for the
brightest magnitude bin plotted in the left panel of Figure
\ref{fig_lambda}; when restricting $N_{mode}$ to 100 or 150, the
fitting occurs on the monotonically rising side of the power spectrum.

The main sources of systematic uncertainty in estimating these
parameters of the 3-dimensional power spectrum from the imaging data
are as follows: (a) the shape of the redshift distribution of the galaxies in a
given magnitude bin; (b) the effects of the cosmological parameters,
primarily the mean mass density, $\Omega_m$, and the vacuum energy
density, $\Omega_\Lambda$, on the redshift-distance relation; (c)
effects of seeing and reddening on the star-galaxy separation in the
data; (d) evolutionary effects, including corrections due to nonlinear
evolution and biasing.

An alternative parameterization of the power spectrum is obtained if
we do not put any evolution into the model power spectra. The
resulting measurements of $\Gamma$ and $\sigma_{8L}$ then represent their
values at the redshift corresponding to the peak contribution in the
projection along the line of sight. For the cosmological model we use,
the peak contribution is at $z\simeq 0.25$, though the weight function
over redshift is quite broad. We obtain $\Gamma=0.183 \pm 0.04$
and $\sigma_{8L}=0.785 \pm 0.053$ for the no-evolution model, in close
agreement with the expected linear amplitude at the peak of the 
redshift distribution.

\subsection{Scaling with changes in the redshift distribution}

The effects of uncertainty in the redshift distribution and the
cosmological parameters are degenerate in their effects on
the shape and amplitude of the power spectrum.  Limber's equation
(eq.~\ref{wtheta}) indicates that the spatial power spectrum derived
from the angular distribution of galaxies depends on the redshift
distribution of galaxies and on the cosmological redshift-distance
relation.  In a flat universe with a non-evolving power spectrum, the
angular power spectrum (and hence its Legendre transform $w(\theta)$)
scales as
\begin{equation}
    C_l = \int {d\chi\over\chi^2} \left(dp\over d\chi\right)^2 P({l/\chi}).
\label{eq:scaling}
\end{equation}
This indicates that if the probability distribution of galaxy
distances $dp/d\chi$ is dilated by a constant $A$, then the inferred
power spectrum will be shifted in wavenumber by a factor $A^{-1}$ and
the power at a given shifted scale will be increased by a factor
$A^3$.  This can be seen intuitively because a dilation of scale of
the universe must scale the wavenumber as an inverse length dilation
and the power spectrum (which has dimensions of volume) as the cube of
the length dilation.

Examining this scaling in more detail, note that if the kernel in
equation (\ref{eq:scaling}) is narrow relative to changes in the power
spectrum, then the amplitude of the spatial power spectrum must scale
as the inverse of
\begin{equation}
    \int {d\chi\over\chi^2} \left(dp\over d\chi\right)^2,
\label{eq:limber}
\end{equation}
whereas the effective wavenumber of the power spectrum sampled by
$C_l$ scales as the inverse of
\begin{equation}
    {\int {d\chi\over\chi^2} \left(dp\over d\chi\right)^2 \chi \over
	\int {d\chi\over\chi^2} \left(dp\over d\chi\right)^2}.
\label{eq:limber2}
\end{equation}
For example, this scaling implies that a 10\% increase in the typical
distance to a galaxy in the sample (i.e.~a 10\% increase in $z_0$ or a
substantial change in the cosmological model) would decrease the
inferred value of $\Gamma$ by 10\% (because the peak scale
$k_{peak}\propto \Gamma$) while increasing the amplitude of the power
spectrum by 30\%.  However, the effects on $\sigma_{8L}$ would be
smaller because the shift of the peak alters the effective slope of
the power spectrum over the range of wavenumbers that contribute most
strongly to $\sigma_{8L}$.  After a 10\% shift, the new power spectrum
will have a value of $\sigma_{8.8}$ equal to the original value of
$\sigma_{8L}$.  Since the value of $\sigma_R$ scales as
$R^{-(n+3)/2}$, where $n$ is an effective spectral index.  At $R\sim
8h^{-1}$Mpc, $n\approx-1.5$, while $n\approx-1$ at larger scales,
where the fluctuations are still linear, thus the value of
$\sigma_{8L}$ scales between $R^{0.75}$ to $R$. Thus, $\Gamma$ and
$\sigma_{8L}$ have an almost inverse relationship. In an excellent
agreement with the above arguments, we find empirically that the
product $\Gamma\sigma_{8L}$ stays approximately constant with respect
to variations in either $dn/dz$ or the underlying cosmology.

\begin{figure}[t!]
\vspace{5.75truein}
\caption{
The effects of various systematics on the likelihood contours for
$\sigma_{8L}$ and $\Gamma$.
{\it Top row:} Varying $dn/dz$ to have median redshifts $z_m=0.293,
0.33, 0.367$, respectively. The middle uses the best fit $dn/dz$,
bracketed by the lower and upper bounds of the estimated distribution
(dashed lines on Figure \ref{fig_sf}. The magnitude range used is
$20<r^*<21$.
{\it Middle row:} Varying the assumed cosmology from our fiducial
model to $\Omega_m=1, \Omega_\Lambda=0.0$.  The three magnitude bins
are as in figure \ref{fig_lambda}.
{\it Bottom row:} Varying the angular extent of the stripe.  We split
our data along the RA direction into two halves, each
2.5$^{\circ}$x44$^{\circ}$. The left and right panels show
$190^{\circ}\le \alpha\le 234^{\circ}$ and $146^{\circ}\le \alpha\le
190^{\circ}$, respectively.  The middle panel shows the full stripe as
in figure \ref{fig_lambda}. The tilting of the likelihood contours in
the half-data stripes shows that reduced coverage in angular scale
leads to a correlation between $\Gamma$ and $\sigma_{8L}$.  The
magnitude range is $20<r^*<21$. Agreement between the two halves, which
have very different average seeing, indicates that seeing does not
affect these results.
}
\includegraphics{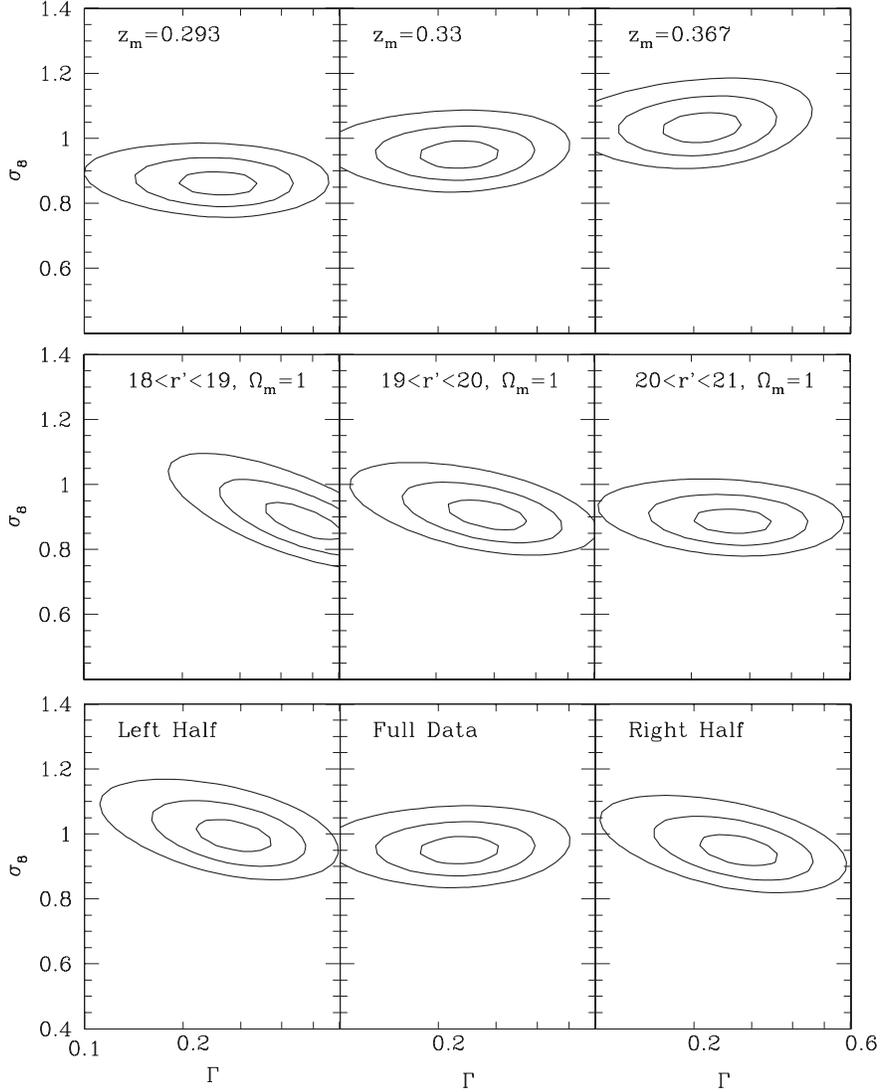}
\label{fig_sys}
\end{figure}

Figure \ref{fig_sys} shows that this scaling relation works remarkably
well.  In this test we compute likelihood contours for $\sigma_{8L}$ and
$\Gamma$ as before, but vary the median redshift of the assumed
redshift distribution for the $20<r^*<21$ sample.  The comoving
distances of the galaxies can also be changed by varying the
cosmological parameters $\Omega_m$ and $\Omega_\Lambda$, which alters
the redshift-distance relation and the evolution of galaxy clustering.
To test the effect of cosmology, in figure \ref{fig_sys} we show
likelihood contours of $\sigma_{8L}$ and $\Gamma$ for the
$20<r^*<21$ sample, this time using $\Omega_m=1,
\Omega_{\Lambda}=0$. This may be compared with figure \ref{fig_lambda}
for the $\Omega_m=0.3, \Omega_{\Lambda}=0.7$ model.  Again, the
predicted scaling of $\sigma_{8L}$ and $\Gamma$ is consistent with the
differences in the parameters obtained for the two models.  We find
that the scaling of $\sigma_{8L}$ and $\Gamma$ in our $20<r^*<21$
sample, is well fit by
\begin{equation}
    \Gamma\sigma_{8L} = 0.173\pm 0.002
\label{eq:fit-scaling}
\end{equation}
in these tests where we vary the redshift distribution or the
cosmological model.

Finally, the evolution of galaxy clustering may differ from the
evolution of matter clustering; in our models we assume that they
evolve identically.  Extant constraints on the amplitude of mass
clustering suggest $\sigma_{8L}({\rm mass})\approx 1$ and we find
$\sigma_{8L}({\rm galaxies})\approx 1$ in these samples, so the
average bias between these galaxies and the underlying mass
distribution is relatively small. Thus, our assumption is roughly
correct for modes in the linear regime, which dominate our results.
However, the observed galaxy clustering amplitude is an average over a
heterogeneous population, whose constituents may undergo different
clustering evolution. Thus, in detail, there may be mild shifts
between the results derived from the different magnitude cuts, which
sample somewhat different populations, not to mention possible
color/morphological type effects.

\subsection{Subsamples of the data set: 
	effects of seeing and angular coverage}

The effects of seeing and galactic extinction are extensively discussed 
by Scranton et al. (2001). With the Bayesian star-galaxy separation
method, these effects are shown to be negligible for the analysis of
galaxy clustering up to the magnitude limit used in this paper.  Here we
perform a further test for the possible effects of variable seeing by
subdividing the data into two halves, one of which suffered from
substantially poorer seeing.  As shown in figure \ref{fig_sys}, we
verify that the power spectrum parameters we obtain are fully
consistent between the two halves, each with different median seeing,
and the full data set.  Note that reducing the area of sky increases
the covariance between $\sigma_{8L}$ and $\Gamma$ because the fits
increasingly depend on wavelength modes that lie to one side of the
peak of the power spectrum, as discussed in section 4.1.

\section{Discussion}


Using only the first 1/50 of the SDSS imaging survey we obtain strong
constraints on the shape and amplitude of the three-dimensional power
spectrum.  Despite lacking redshifts, this photometric sample covering
222 sq. degrees yields uncertainties on $\Gamma$ that are only
slightly larger than those estimated from the 2dFGRS sample of 160,000
spectroscopic redshifts (Percival et al. 2001), the largest redshift
survey to date. For the faintest apparent magnitude subsample that we
examine, $20<r^*<21$, the fitted parameters of the
power spectrum extrapolated to $z=0$ are $\Gamma=0.188\pm 0.04$ and
$\sigma_{8L}=0.915\pm 0.06$ ($1\sigma$ uncertainties from marginalizing
over one parameter at a time). We find a trend toward larger $\Gamma$
in our brightest subsample, which reaches into mildly nonlinear
scales, biasing the $\Gamma$ estimate.  Thus, the $20<r^*<21$ sample
yields the best estimate of the linear power spectrum because it
probes the largest angular scales and is not affected by nonlinearity.


The ability to quickly identify modes that are useful for linear power
spectrum estimation is a strong advantage of the KL analysis method.
For our stripe-like sky geometry, the eigenmodes naturally segregate
into modes that probe large wavelengths along the stripe, short
wavelengths across the stripe, and mixtures of the two.  Examination
of the range of Fourier modes probed by the KL eigenmodes of the
sample shows that the highest signal-to-noise KL modes are sensitive
primarily to large wavelength fluctuations that lie along the right
ascension axis of the stripe.  We find that these modes probe scales beyond
the peak of the best-fit CDM-like power spectrum. In other words, the
KL modes make optimal use of the widest direction of our sample area. For
the deepest sample we examine, $20<r^*<21$ which has
$z_m=0.33$, the peak sensitivities of the ten highest signal-to-noise
modes are all at comoving wavelength $2\pi/k > 200h^{-1}$Mpc.  We
examine the sensitivity of the fitted parameters to the number of
modes used in the analysis, plot the wavenumbers probed by these modes
and, as we expect, find that the fitting is stable when nonlinear
modes are excluded.


Various estimates of the power spectrum from the SDSS EDR-P sample are
also provided by Connolly et al. (2001), Dodelson et al. (2001), Tegmark
et al. (2001), and from a galaxy redshift sample over a similar region
of the SDSS by Zehavi et al. (2001).  In all of the analyses of the SDSS
EDR-P photometric sample, we find that the fitted parameters depend on
how we choose to limit the range of wavelength scales used in the
fitting procedure. Inclusion of nonlinear modes tends to raise
$\Gamma$ and lower $\sigma_{8L}$ (see section 4.1). Thus, small
variations among the fitted parameters arise when using different
estimation methods because they use different projections of the data
(KL eigenmodes, spherical harmonics, angular pair counts), which vary
in the manner in which they segregate power at linear vs. nonlinear
scales. The ease with which one can examine the same range of scales
depends on the method. 


For this first assay of the KL method on a photometric sample, we
choose to limit the estimated parameters to the shape, represented by
$\Gamma$ and extrapolated linear amplitude $\sigma_{8L}$.  Larger
galaxy samples and CMB data can probe the matter density parameter
$\Omega_m h^2$ and baryon to total matter ratio $\Omega_b/\Omega_m$
separately rather than $\Gamma$.  Percival et al. (2001) argue that
the 2dFGRS is large enough to do so.  They obtain $\Omega_m h =
0.20\pm 0.03$ and $\Omega_b/\Omega_m=0.15\pm 0.07$ for a redshift sample
of 160,000 galaxies to $b_J=19.45$.  We can use the approximate
formula of Sugiyama (1995) to convert this to the shape parameter,
$\Gamma = \Omega_m h/ \exp(\Omega_b (1+\sqrt{2h}/\Omega_m))$. For the
estimates of Percival et al., this yields $\Gamma=0.17\pm 0.03$, with
which we agree within 1$\sigma$ for our best (deepest) sample.

It is also instructive to compare with parameters estimated from
recent CMB anisotropy experiments. Results from DASI and Boomerang
(Pryke et al. 2001, Netterfield et al. 2001) correspond to
\begin{equation}
        0.16 < \Omega_m h < 0.27; \qquad 0.10 < \Omega_b/ \Omega_m < 0.18.
\label{eq:cmb}
\end{equation}
These bounds are based on combining their estimates for the dark
matter, baryonic matter and errors with a strong Hubble prior of
$h=0.72\pm 0.08$ (HST Key Project -- Freedman et al. 2001), and
assuming the distributions are disjoint and normally distributed with
the quoted errors.

\begin{figure}[t!]
\vspace{6truein}
\caption{ 
A plot of the CMB constraints on $\Omega_b/\Omega_m$ vs $\Omega_m h$,
and $\Gamma$ vs $\sigma_{8L}$.  The points were generated by $10^6$
Monte Carlo simulations, assuming gaussian distributions for $\Omega_c
h^2$, $\Omega_b h^2$, and $n$ from DASI (Pryke et al 2001) and
BOOMERANG (Netterfield et al 2001), as well as $h=0.72\pm0.08$. The
upper plot shows the 68\% confidence region in the $\Omega_b/\Omega_m$
vs $\Omega_m h$ plane, and the bottom plot shows the 68\% confidence
region in the $\sigma_{8L}$ vs $\Gamma$ plane. Our fiducial contour is
marked as the ellipse on the lower plot.  The plots use dark x's to
mark those cells for which at least 25\% of the models fall in the
$\Gamma-\sigma_{8L}$ error ellipse of our paper. These dark x's may
thus be regarded as the set of model parameters jointly allowed by the
CMB and our LSS constraints. Values as high as $\Gamma=0.24$ and as
low as $\Gamma=0.15$ are possible.  The points on the upper plot form
a smooth band; the CMB and LSS constraints are almost orthogonal.
}
\includegraphics{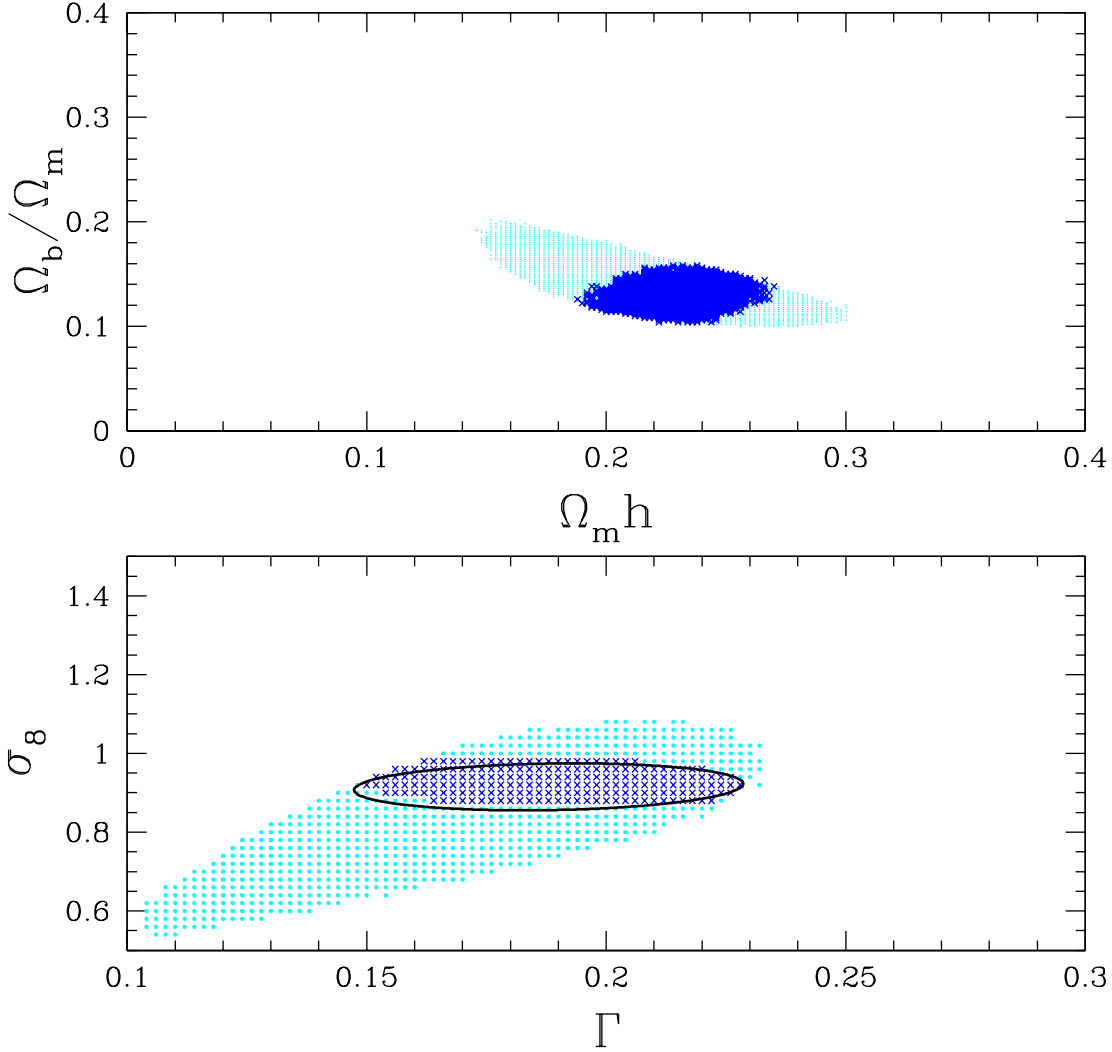}
\label{fig_cmb}
\end{figure}

A plot of the CMB constraints on $\Omega_b/\Omega_m$ vs $\Omega_m h$,
and $\Gamma$ vs $\sigma_{8L}$ is shown in Figure \ref{fig_cmb}.  The
points were generated by $10^6$ Monte Carlo simulations, assuming
Gaussian distributions for $\Omega_c h^2$, $\Omega_b h^2$, and $n$
from DASI (Pryke et al 2001) and BOOMERANG (Netterfield et al 2001),
as well as $h=0.72\pm0.08$. The upper plot shows the 68\% confidence
region in the $\Omega_b/\Omega_M$ vs $\Omega_M h$ plane, and the
bottom plot shows the 68\% confidence region in the $\sigma_{8L}$ vs
$\Gamma$ plane. Our fiducial contour is marked as the ellipse on the
lower plot.  The plots use dark x's to mark those cells for which at
least 25\% of the models fall in the $\Gamma-\sigma_{8L}$ error
ellipse of our paper. These dark x's may thus be regarded as the set
of model parameters jointly allowed by the CMB and our LSS
constraints. Values as high as $\Gamma=0.24$ and as low as
$\Gamma=0.15$ are possible.  The points on the upper plot form a
smooth band; the CMB and LSS constraints are almost orthogonal. The
values for $\Omega_b/\Omega_M$ are in excellent agreement with the
most recent results from Big Bang Nucleosynthesis (Burles et al 2001).


We also examine the evolution of clustering by estimating parameters
of the galaxy power spectrum at the effective redshift of each
subsample (rather than extrapolating that clustering to zero redshift
as in the previous analyses).  Note that different ranges of absolute
magnitude are sampled by the different apparent magnitude slices, thus
luminosity dependence of clustering complicates interpretation of
these results; the signal of genuine clustering evolution remains to
be disentangled from the systematic effect of varying the intrinsic
luminosity of galaxies in the three apparent-magnitude limited
subsamples that we examine.  


This analysis has only used about 2\% of the whole SDSS data set. It
is clear that the statistical accuracy is going to improve
dramatically for the whole data set. Systematic uncertainties, like
photometric calibrations and extinction corrections will be the
limiting factor at that point, though these are also going to improve
by factors of several. Photometric redshifts (Connolly et al 1995)
offer an elegant extension of this method: by selecting several thick
photo-z slices we can measure the shape parameters of the power
spectrum at several redshifts, thus measuring evolution in the
clustering. By deriving an SED type for each galaxy we can also create
rest-frame selected samples at different redshifts. This paper has
shown that even without redshifts, but with accurate photometry one
can derive surprisingly accurate information about the shape of the
primordial fluctuations.

\acknowledgements

The Sloan Digital Sky Survey (SDSS) is a joint project of The
University of Chicago, Fermilab, the Institute for Advanced Study, the
Japan Participation Group, The Johns Hopkins University, the
Max-Planck-Institute for Astronomy (MPIA), the Max-Planck-Institute
for Astrophysics (MPA), New Mexico State University, Princeton
University, the United States Naval Observatory, and the University of
Washington. Apache Point Observatory, site of the SDSS telescopes, is
operated by the Astrophysical Research Consortium (ARC).  Funding for
the project has been provided by the Alfred P. Sloan Foundation, the
SDSS member institutions, the National Aeronautics and Space
Administration, the National Science Foundation, the U.S. Department
of Energy, the Japanese Monbukagakusho, and the Max Planck
Society. The SDSS Web site is http://www.sdss.org/.

AS has been supported by NSF AST-9802980 and NASA NAG5-53503.
BJ acknowledges support from NASA through grants NAG5-9186 and
NAG5-9220.  MSV acknowledges support from NSF grant AST-0071201.


\end{document}